\documentclass[amsmath,amssymb,aps,twocolumn,nofootinbib]{revtex4-2}
\let\svthefootnote\thefootnote
\newcommand\freefootnote[1]{%
  \let\thefootnote\relax%
  \footnotetext{#1}%
  \let\thefootnote\svthefootnote%
}

\usepackage{graphicx}
\usepackage{dcolumn}
\usepackage{bm}
\usepackage[colorlinks]{hyperref}

\begin{document}

\title{Geometric phases and the Sagnac effect: Foundational aspects and sensing applications}

\author{I. L. Paiva}
\affiliation{Faculty of Engineering and the Institute of Nanotechnology and Advanced Materials, Bar-Ilan University, Ramat Gan 5290002, Israel}

\author{R. Lenny}
\affiliation{Faculty of Engineering and the Institute of Nanotechnology and Advanced Materials, Bar-Ilan University, Ramat Gan 5290002, Israel}

\author{E. Cohen}
\affiliation{Faculty of Engineering and the Institute of Nanotechnology and Advanced Materials, Bar-Ilan University, Ramat Gan 5290002, Israel}

\begin{abstract}
Geometric phase is a key player in many areas of quantum science and technology. In this review article, we outline several foundational aspects of quantum geometric phases and their relations to classical geometric phases. We then discuss how the Aharonov-Bohm and Sagnac effects fit into this context. Moreover, we present a concise overview of technological applications of the latter, with special emphasis on gravitational sensing, like in gyroscopes and gravitational wave detectors.
\end{abstract}

\maketitle

\section{Introduction}

\freefootnote{This is the peer reviewed version of the following article: Adv. Quantum Technol. 5, 2100121 (2022), which has been published in final form at \url{https://doi.org/10.1002/qute.202100121}. This article may be used for non-commercial purposes in accordance with Wiley Terms and Conditions for Use of Self-Archived Versions. This article may not be enhanced, enriched or otherwise transformed into a derivative work, without express permission from Wiley or by statutory rights under applicable legislation. Copyright notices must not be removed, obscured or modified. The article must be linked to Wiley’s version of record on Wiley Online Library and any embedding, framing or otherwise making available the article or pages thereof by third parties from platforms, services and websites other than Wiley Online Library must be prohibited.}

The fundamental notion of geometric phase \cite{berry1984quantal,berry1988geometric,shapere1989geometric} has been highly influential in physics and related sciences. It lies at the heart of many quantum phenomena pertaining to basic science and at the same time delineates important technological applications for quantum sensing and quantum computation.

Various geometric phases were originally introduced as a result of physical systems undergoing adiabatic cyclic evolution. However, this notion was extended with the removal of the adiabatic and even the cyclic conditions. In this work, we intend to review fundamental aspects underlying these phases and present some of their practical applications, such as in gyroscopes and Laser Interferometer Gravitational-Wave Observatory (LIGO) detectors.

Differently from other review articles on the subject, see, e.g., Refs. \cite{olariu1985quantum, zwanziger1990berry,anandan1992geometric, cohen2019geometric, karnieli2022geometric}, we focus on the relation between geometric phases and accelerating as well as gravitational systems. In particular, we give special emphasis on quantum applications of the Sagnac effect in such systems. At the same time, much more attention is given to geometric phases here than in review articles specialized on Sagnac interferometers, such as Refs. \cite{post1967sagnac, anderson1994sagnac, schreiber2013invited}.

The paper is organized as follows. Section \ref{sec:geometry} discusses the geometry underlying quantum systems. From this structure, geometric phases are introduced in Section \ref{sec:geom-phases}. This section also compares quantum geometric phases with classical ones. The Aharonov-Bohm (AB) effect, which is associated with a special --- i.e., topological --- type of geometric phase, is presented in Section \ref{sec:ab-effect}. Following that, Section \ref{sec:sagnac-theory} discusses geometric phases in non-inertial frames and introduces the Sagnac effect. Then, Section \ref{sec:sagnac-application} presents various applications of the Sagnac effect for quantum sensing and metrology. Section \ref{sec:ab-gravitational} presents geometric and gravitational AB-like effect and discuss their relevance in the present context. Finally, Section \ref{sec:discussion} concludes the paper with some future outlook.

\section{Geometry of quantum states}
\label{sec:geometry}

In this section, we present the geometry underlying the mathematical structures of quantum states. Although the formal name of each structure is mentioned, we restrict our presentation to the intuition behind them. For a more technical exposition, we refer the reader to, e.g., Refs. \cite{marsden1990reduction, chruscinski2004geometric}.

Consider a pure quantum system represented by a state in a complex Hilbert space $\mathcal{H}_{n+1}\setminus\mathcal{O}$, where $n\in\mathbb{N}$, $n+1$ is the dimension of the space, and $\mathcal{O}$. Because of the probabilistic interpretation, it is conventional to work with normalized vectors. Also, states that differ only by a global phase are indistinguishable. Then, two non-null vectors $|\psi\rangle$ and $|\varphi\rangle$ in $\mathcal{H}_{n+1}$ represent the same physical state if there exists $\lambda\in\mathbb{C}$ such that
\begin{equation}
    |\psi\rangle = \lambda |\varphi\rangle.
    \label{eq:relation-cp}
\end{equation}
Then, while $n+1$ complex parameters are needed to characterize a vector in $\mathcal{H}_{n+1}$, $n$ suffices to characterize a pure quantum state.

Eq. \eqref{eq:relation-cp} effectively defines an equivalence relation in $\mathcal{H}_{n+1}$, subdividing the space into \textit{equivalence classes}. In fact, denoting this relation by $\sim$, we can write
\begin{equation}
    |\psi\rangle \sim |\varphi\rangle
\end{equation}
whenever there exists $\lambda\in\mathbb{C}$ such that Eq. \eqref{eq:relation-cp} is satisfied. Here, again, $|\psi\rangle$ and $|\varphi\rangle$ are non-null vectors in $\mathcal{H}_{n+1}$.

The resultant space after the operation $\sim$ is applied to entire $\mathcal{H}_{n+1}\setminus\mathcal{O}$ is the \textit{projective space}, also known as \textit{ray space},
\begin{equation}
    \mathbb{CP}(n)\equiv \frac{\mathcal{H}_{n+1}\setminus\mathcal{O}}{\sim},
\end{equation}
which is an $n$-dimensional complex space.

\begin{figure*}
    \centering
    \includegraphics[width=\textwidth]{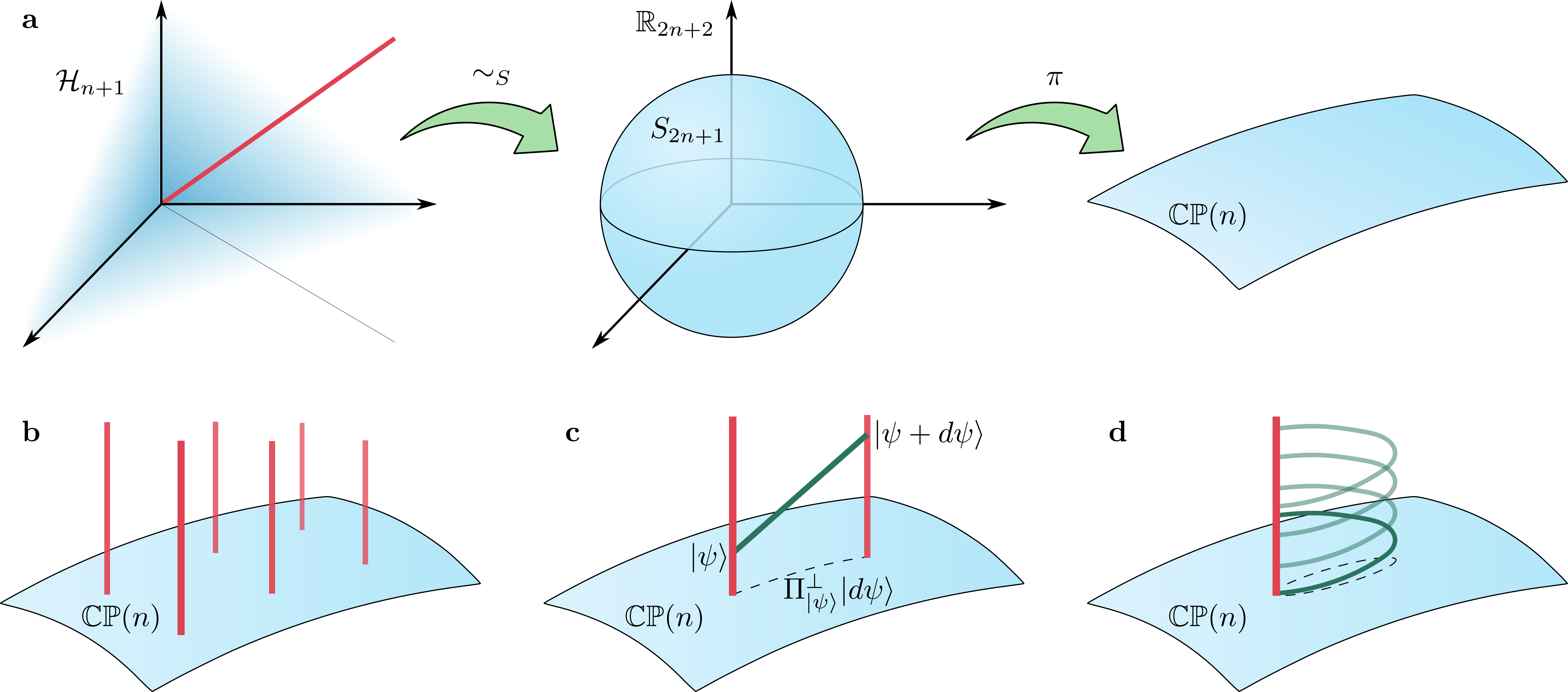}
    \caption{Geometric representation of pure quantum systems whose state is an element of a complex Hilbert space $\mathcal{H}_{n+1}$, which is isomorphic to a real vector space of dimension $2n+2$. \textbf{(a)} An arbitrary direction of the Hilbert space, like the red line, represents a single physical state. Then, restricting to normalized vectors with an equivalence relation $\sim_S$, the resulting space is isomorphic to a real hypersphere $S_{2n+1}$. Finally, a projector $\pi$ can map each point that differs by a global phase to a single element of a complex projective space $\mathbb{CP}(n)$. \textbf{(b)} The global phases, thus, can be represented by tangent fibers to each point of $\mathbb{CP}(n)$. \textbf{(c)} Using the fiber bundle structure, a notion of distance can be naturally introduced in $\mathbb{CP}(n)$ since the projector of orthogonal direction to a certain state $|\psi\rangle$, represented by $\Pi_{|\psi\rangle}^\perp$ is well-defined. \textbf{(d)} Moreover, a system that completes a loop in $\mathbb{CP}(n)$ does not always return to its initial point in the bundle structure. This lack of holonomy constitutes the geometric origins of geometric phases.}
    \label{fig:geometry}
\end{figure*}

It is typically convenient to treat $\sim$ as a composition of two distinct equivalence relations. One of them, denoted by $\sim_S$ is such that any two vectors $|\psi\rangle$ and $|\varphi\rangle$ vectors in $\mathcal{H}_{n+1}\setminus\mathcal{O}$ are equivalent, i.e.,
\begin{equation}
    |\psi\rangle \sim_S |\varphi\rangle,
\end{equation}
whenever there exists $\rho\in\mathbb{R}$ such that
\begin{equation}
    |\psi\rangle = \rho |\varphi\rangle.
\end{equation}
It can be checked that the resultant space is isomorphic to the unity $2n+1$-dimensional real hyper-sphere $S_{2n+1}$, as represented in Figure \ref{fig:geometry}a. In fact, with the relation $\sim_S$, the $n+1$ complex parameters (or $2n+2$ real ones) necessary to characterize a state in $\mathcal{H}_{n+1}$ are constrained by an equation that establishes the unit norm, which results in $2n+1$ free real parameters necessary for the characterization of a state. Because of this isomorphism, the space associated with $\sim_S$ will be referred to as the hypersphere $S_{2n+1}$.

To complete the relation $\sim$, a map $\pi$, also associated with an equivalence relation, is applied to the elements of $S_{2n+1}$, mapping them into a representative element of their class in $\mathbb{CP}(n)$, as illustrated in Figure \ref{fig:geometry}a. Observe that if two points associated with the normalized vectors $|\psi\rangle$ and $|\varphi\rangle$ are such that
\begin{equation}
    \pi(|\psi\rangle) = |\varphi\rangle,
\end{equation}
then there exists $\theta\in\mathbb{R}$ such that
\begin{equation}
    |\psi\rangle = e^{i\theta} |\varphi\rangle,
\end{equation}
i.e., they differ by a global phase. Hence, these phases can be seen as fibers with circular topology at each point of $\mathbb{CP}(n)$, as represented in Figure \ref{fig:geometry}b. In fact, this constitutes a \textit{fiber bundle} structure, where $\mathbb{CP}(n)$ is the base space, the space of the phases is the fiber, $S_{2n+1}$ is the total space, and $\pi$ is the projection map, also known as bundle projection.

It can be shown that the space $\mathbb{CP}(n)$ inherits the symplectic structure of $\mathcal{H}_{n+1}$, i.e., a well-defined notion of hypervolume \cite{chruscinski2004geometric}. Thus, the projection of the temporal evolution of the unitary dynamics on $\mathbb{CP}(n)$ is a Hamiltonian dynamics.

In addition to this structure, $\mathbb{CP}(n)$ also acquires the metric induced by the inner product in $\mathcal{H}_{n+1}$. It is, then, possible to compute the distance between two points in $\mathbb{CP}(n)$. In fact, let $|\psi\rangle$ and $|\psi+d\phi\rangle$ be infinitesimally close in $S_{2n+1}$. Then, as illustrated in Figure \ref{fig:geometry}c, a natural definition for the distance in $\mathbb{CP}(n)$ between them is
\begin{equation}
    ds^2(\mathbb{CP}(n)) \equiv K \langle d\psi| \Pi_{|\psi\rangle}^\perp |d\psi\rangle,
    \label{eq:metric-def}
\end{equation}
where $K$ is a positive real constant, $\Pi_{|\psi\rangle}^\perp \equiv I-|\psi\rangle\langle\psi|$ is the projector in the orthogonal direction to $|\psi\rangle$, and $|d\psi\rangle = |\psi+d\psi\rangle - |\psi\rangle$. Interestingly, if $|\psi+d\psi\rangle$ is such that $|\psi+d\psi\rangle=|\psi(t+dt)\rangle$, i.e., is the result of an evolution in time generated by a Hamiltonian $H$, we obtain
\begin{equation}
    |d\psi(t)\rangle = |\psi(t+dt)\rangle - |\psi(t)\rangle = -\frac{i}{\hbar} H |\psi(t)\rangle dt
\end{equation}
from the Schr\"odinger equation. Then, the distance between these two states in $\mathbb{CP}(n)$ is
\begin{equation}
    \begin{aligned}
        ds^2(\mathbb{CP}(n)) &= K \left(\langle d\psi|d\psi\rangle -  \langle d\psi|\psi\rangle\langle\psi|d\psi\rangle\right) \\
            &= \frac{K}{\hbar^2} \left(\langle \psi|H^2|\psi\rangle -  \langle\psi|H|\psi\rangle^2\right) dt^2 \\
            &= \frac{K}{\hbar^2} \left(\Delta H\right)^2 dt^2,
    \end{aligned}
\end{equation}
i.e., the ``velocity'' of the system in $\mathbb{CP}(n)$ is proportional to the uncertainty of its energy, as shown by Anandan and Aharonov \cite{anandan1990geometry}.

Interestingly, it follows from the Schr\"odinger equation that the distance between two states $|\psi(t)\rangle$ and $|\varphi(t)\rangle$ evolving under the same Hamiltonian do not change in time since
\begin{equation}
    \frac{d}{dt} \left(\langle\psi(t)|\varphi(t)\rangle\right) = 0
    \label{eq:const-prob-trans}
\end{equation}
and Eq. \eqref{eq:metric-def} implies that, in $\mathbb{CP}(n)$, the distance $s(|\psi\rangle,|\varphi\rangle)$ between two states $|\psi\rangle$ and $|\varphi\rangle$ is such that
\begin{equation}
    s^2(|\psi\rangle,|\varphi\rangle) = K \left(1-|\langle\psi|\varphi\rangle|^2\right).
\end{equation}
Physically, this means that the probability of transition between these two states is also preserved. In particular, if $|\psi(t)\rangle = |\varphi(t)\rangle$, it follows that the norm of vectors is kept constant in time. In other words, the evolution of a state creates a trajectory in $S_{2n+1}$, which can, then, be projected into $\mathbb{CP}(n)$.

Another important aspect is that, although the global phase of a state vector has no physical meaning, the phase difference between two states has. It is even possible to define a \textit{connection} in $S_{2n+1}$ by establishing that, given two states $|\psi\rangle$ and $|\psi+d\psi\rangle$ projected into points infinitesimally close to each other in $\mathbb{CP}(n)$, they are parallel to each other in $S_{2n+1}$ if the phase difference between them is null, i.e.,
\begin{equation}
    \arg\left(\langle\psi|\psi+d\psi\rangle\right) = 0.
\end{equation}

With the connection, the concept of \textit{parallel transport} can be introduced. Given a curve in $\mathbb{CP}(n)$ represents an infinite number of curves in $S_{2n+1}$. A subset of special interest is the one for which the curves are defined on geodesics in $S_{2n+1}$, i.e., the curves built through parallel transport in $S_{2n+1}$. These trajectories are characterized by
\begin{equation}
    \langle\psi(s)| \frac{d}{ds} |\psi(s)\rangle = 0
\end{equation}
along the curve defined by $|\psi(s)\rangle$ in terms of a real parameter $s$. Each of these curves is called a \textit{geodesic lifting} of the curve in $\mathbb{CP}(n)$. A geodesic lifting does not have intrinsic physical meaning --- only the family of geodesic liftings has. This is illustrated in Figure \ref{fig:geometry}d, where all geodesic liftings (green curves) have an equivalent physical meaning.

Besides its significance to the study of geometric phases, as it will be discussed in the next section, the geometric structure presented here was used for a pedagogical approach to the quantum adiabatic theorem \cite{lobo2012geometry} and signaling in Weinberg's non-linear quantum mechanics \cite{paiva2014alguns}. Also, this structure provides a geometric interpretation to von Neumann's measurement interaction model, weak values, and quantum erasers \cite{tamate2009geometrical, lobo2014weak}, and can be used for the introduction of a type of time-energy uncertainty relation \cite{anandan1990geometry}. Finally, this structure is also the same that appears in Yang-Mills theories, which underlie the standard model of particle physics.

\section{Geometric phases}
\label{sec:geom-phases}

As a consequence of the geometry discussed in the previous section, when a system completes a closed cycle in $\mathbb{CP}(n)$, its trajectory is not necessarily closed in $S_{2n+1}$. A measure of this lack of \textit{holonomy} can be locally defined at each point of $\mathbb{CP}(n)$ by considering an infinitesimal closed curve around each point. This quantity corresponds to the \textit{curvature} of the connection and is the mathematical structure in which we are interested. In fact, this structure stands behind the geometric (or non-integrable) phase, as represented in Figure \ref{fig:geometry}d.

To see that, assume a system, initially in the state $|\psi(0)\rangle$, completes a cyclic evolution of period $\tau$ in $\mathbb{CP}(n)$. Then, we can write its final state as
\begin{equation}
    |\psi(\tau)\rangle = e^{i\phi} |\psi(0)\rangle.
\end{equation}
Part of the phase it acquires is the dynamical phase
\begin{equation}
    \phi_\text{dyn}(\tau) = -\frac{1}{\hbar} \int_0^\tau \langle\psi(t)|H(t)|\psi(t)\rangle \ dt,
\end{equation}
where $H(t)$ is the Hamiltonian that governs the dynamics of the system. However, Berry noted that, in the adiabatic regime, the system in fact accumulates an extra phase, which became known as the \textit{Berry phase} \cite{berry1984quantal}. This extra phase is geometric and corresponds to the lack of holonomy we just discussed, as it was shown by Simon \cite{simon1983holonomy}. Because of it, the natural connection introduced above is known as the Berry-Simon connection. Later on, the adiabatic condition was removed by Aharonov and Anandan \cite{aharonov1987phase}. In fact, defining
\begin{equation}
    |\tilde{\psi}(t)\rangle \equiv e^{-i \left[f(t) + \phi_\text{dyn}(t)\right]} |\psi(t)\rangle,
\end{equation}
where $f$ is a function such that $\phi-\phi_\text{dyn}(\tau)=f(\tau)-f(0)$, it follows that $|\tilde{\psi}(\tau)\rangle=|\tilde{\psi}(0)\rangle$ and
\begin{equation}
    i\hbar \frac{d}{dt} |\tilde{\psi}(t)\rangle = \left[ H(t) + \hbar \frac{d}{dt} f(t) - \langle\psi(t)|H(t)|\psi(t)\rangle \right] |\tilde{\psi}(t)\rangle,
\end{equation}
which, in turn, implies that
\begin{equation}
    \frac{d}{dt}f(t) = \langle \tilde{\psi}(t) | \left(i \frac{d}{dt} \right) | \tilde{\psi}(t) \rangle,
\end{equation}
regardless of $H$. Then, after ending the cycle, the state $|\psi\rangle$ accumulates an extra phase $\phi_\text{geom} \equiv f(\tau) - f(0)$ that can be expressed as
\begin{equation}
    \phi_\text{geom} = \int_0^\tau \langle \tilde{\psi}(t) | \left(i \frac{d}{dt} \right) | \tilde{\psi}(t) \rangle \ dt = i \oint_C \langle \tilde{\psi}(R) | d\tilde{\psi}(R) \rangle,
    \label{def-geom-phase}
\end{equation}
which can be shown to depend only on geometric properties associated with the cycle, thus, in particular, is not a function of $\tau$. The last integral in Eq. \eqref{def-geom-phase} is taken over a curve $C$ in the parameter space, which has a parameter $R$ associated with it. These phases also appear in adiabatic quantum transitions \cite{berry1990geometric}. Moreover, they are present and play a fundamental role in the properties of many physical systems in nature, like in molecular systems \cite{mead1992geometric} and crystalline dielectrics \cite{resta1994macroscopic}.

Furthermore, geometric phases can be introduced for mixed states \cite{sjoqvist2000geometric, singh2003geometric}. In fact, they are studied in open and non-Hermitian systems, although there is no unique approach for that \cite{garrison1988complex, zwanziger1991measuring, ning1992geometrical, bliokh1999appearance, filipp2003off, carollo2003geometric, tong2004kinematic, lombardo2006geometric, dietz2011exceptional}. However, a noteworthy remark is that, depending on how it is extended to the dynamics of non-Hermitian Hamiltonians, the phase introduced by Aharonov and Anandan always assumes real values. In this case, the adiabatic limit of $\phi_\text{geom}$ does not always correspond to the Berry phase, which can be complex-valued \cite{wu1996berry}. Moreover, in general scenarios, these phases can be studied in systems with non-cyclic evolution and even when intermediate measurements are taken into account \cite{samuel1988general}.

Also, geometric phases can be introduced for classical systems undergoing adiabatic evolution, as it was shown by Hannay \cite{hannay1985angle}. In fact, for integrable systems, one can represent the Hamiltonian that governs the evolution in terms of action-angle coordinates \cite{arnol1989mathematical}. It can be, then, concluded that the evolution is defined in a torus in phase space. In cyclic adiabatic evolutions, action coordinates are preserved while angle coordinates may change. Hannay observed that these changes in angle coordinates during a cycle have dynamical and geometrical contributions, in a similar manner first observed by Berry for quantum systems. The geometrical contribution is what became known as the Hannay angle. From a more geometrical perspective, the Cartesian product between the parameter space and the phase-space defined a trivial fiber over the parameter space. Within this structure, the Hannay angle is, then, obtained from liftings in the total space according to a connection known as the Hannay-Berry connection \cite{golin1989hannay, marsden1990reduction, chruscinski2004geometric}. Then, like for Berry phases, the notion of a Hannay angle for open paths can also be considered \cite{pati1998adiabatic}.

It should be noted that the notion of geometric phases can be extended to classical dissipative systems \cite{kepler1991geometric}. Moreover, perturbative methods can be used for nonadiabatic correction in some scenarios \cite{andersson2005nonadiabatic}. However, the corrections are not purely geometric since, in general, they depend on the time-parametrization of the trajectories. Furthermore, geometric phases are also studied in nonlinear classical field theories \cite{garrison1988geometrical, anandan1988comment, latmiral2020berry}.

Given a Hamiltonian of an integrable system and denoting the action variable by $I=\{I_0,\cdots, I_{n-1}\}$, it is possible to find a direct mathematical relation between the Hannay angle $h(I;C)$ and the Berry phase $\phi(C)$ acquired by a physical system during a cyclic evolution through a path $C$. This follows from the use of the Bohr-Sommerfeld quantization rule, in which each $I_j$ is quantized as $I_j = \hbar (n_j + \mu_j/4)$, where $\mu_j$ is a \textit{Maslov index}, which is a topological quantity, and $n_j$ is an integer. This, combined with further analysis of the geometrical structure permeating both phases, leads to \cite{berry1985classical, chruscinski2004geometric}
\begin{equation}
    \frac{\partial\phi(C)}{\partial n_j} = -h(I;C) + O(\hbar).
\end{equation}
While this implies that the Hannay angle may vanish in scenarios where the Berry phase differs from zero, the converse does not hold: the Berry phase cannot vanish if the Hannay angle is null.

This type of relation makes the ability to discriminate quantum and classical contributions to interference experiments an important issue, which was already considered, e.g., in optomechanical systems \cite{armata2016quantum}. Also, it can be argued that geometric phases observed in various experiments with light-waves --- not matter-waves --- are Hannay angles \cite{agarwal1990berry}. This is the case because the Berry phase becomes proportional to the Hannay angle. In this regard, the use of squeezing Hamiltonians was suggested as a possible way to make Berry phases detectable in these experimental setups \cite{chaturvedi1987berry, chiao1988lorentz, fuentes2000proposal}.

Before concluding this section, we call attention to the fact that, in cases of Hamiltonians with degeneracy, geometric phases (both quantum and classical) become non-Abelian \cite{wilczek1984appearance, anandan1988non}.

\section{Aharonov-Bohm effect}
\label{sec:ab-effect}

An important type of geometric phase is the phase associated with the AB effect. In classical physics, the dynamics of a particle with charge $q$ is only affected by a magnetic field that directly interacts with it, i.e., if the particle travels in a region where the magnetic field is non-zero. However, in quantum mechanics, this is not always the case. In fact, suppose a charge encircles a region in space that contains a magnetic field. Then, it accumulates a quantum phase proportional to the magnetic flux inside the region enclosed by its trajectory, regardless of whether there was any magnetic field on the particle's trajectory.

In 1939, this effect might have been hinted at by Franz in a talk at a physical society meeting in Danzig \cite{franz1939}. Later, in 1949, the effect was presented by Ehrenberg and Siday \cite{ehrenberg1949refractive}, although the influence of magnetic field in the particle's dynamics seemed to be a peculiar feature of the optical configuration considered by them. The authors themselves wrote: ``One might therefore expect wave-optical phenomena to arise which are due to the presence of a magnetic field but not due to the magnetic field itself, i.e., which arise whilst the rays are in field-free regions of space.'' It was only in 1959 that the fundamental nature of the effect was revealed in a seminal article by Aharonov and Bohm \cite{Aharonov1959}, the reason for which it is known as the AB effect. More details on early historical aspects of the discovery of the effect can be found in Ref. \cite{hiley2013early}. Since Aharonov and Bohm's work, the AB effect has been vastly investigated in theoretical and experimental works \cite{chambers1960shift, mollenstedt1962kontinuierliche, liebowitz1965significance, aharonov1969modular, boyer1973classical, berry1980wavefront, tonomura1982observation, tonomura1986evidence, berry1986aharonov, berry1986statistics, osakabe1986experimental, berry1989quantum, ford1994aharonov, aharonov1994, aharonov2004effect, aharonov2005quantum, tonomura2006aharonov, recher2007aharonov, russo2008observation, peng2010aharonov, berry2010semifluxon, fang2012photonic, bardarson2013quantum, noguchi2014aharonov, duca2015aharonov, kang2015locality, cohen2015measure, aharonov2016nonlocality, mukherjee2018experimental, paiva2019topological, paiva2020magnetic}.

This effect is usually presented by considering a charge encircling a solenoid whose axis lies, say, on the $z$ axis. For simplicity, the solenoid is taken to be infinitely thin and is sometimes referred to as a \textit{flux line}. Also, for simplicity, the particle is assumed to travel in the $xy$ plane, having the superposition state
\begin{equation}
    |\Psi\rangle = \frac{1}{\sqrt{2}} \left(|\psi_L\rangle + |\psi_R\rangle\right),
\end{equation}
where $|\psi_L\rangle$ is a wavepacket that passes to the left and $|\psi_R\rangle$ is a wavepacket that passes to the right of the flux line. This superposition can be achieved, for instance, with the use of a double-slit or a beamsplitter. The Hamiltonian of the particle, in this case, can be written as
\begin{equation}
    H = \frac{1}{2m} \left[\vec{P}- q\vec{A}(\vec{Q})\right]^2,
    \label{eq-ham-vp}
\end{equation}
where $\vec{P}=P_x\hat{x}+P_y\hat{y}$, $\vec{Q}=X\hat{x}+Y\hat{y}$, and $\vec{A}$ is the vector potential associated with the solenoid, which can take different forms according to the choice of gauge.

To obtain the solution of this Hamiltonian, let the states $|\psi_L^0\rangle$ and $|\psi_R^0\rangle$ be the solutions in the case where there is no magnetic field, i.e., $H_0 = P^2/2m$. Then, based on a procedure introduced by Dirac \cite{dirac1931quantised}, it can be obtained that, after the left and the right wavepackets travel, respectively, the trajectories $\gamma_L$ and $\gamma_R$,
\begin{equation}
    |\psi_L\rangle = e^{iq\int_{\gamma_L}\vec{A}\cdot d\vec{\ell}/\hbar}|\psi_L^0\rangle
    \label{eq-phase-l}
\end{equation}
and
\begin{equation}
    |\psi_R\rangle = e^{iq\int_{\gamma_R}\vec{A}\cdot d\vec{\ell}/\hbar}|\psi_R^0\rangle.
    \label{eq-phase-r}
\end{equation}
Now, recall that quantum states are equivalent up to a global phase and observe that
\begin{equation}
    \int_{\gamma_R}\vec{A}\cdot d\vec{\ell} - \int_{\gamma_L}\vec{A}\cdot d\vec{\ell} = \oint \vec{A}\cdot d\vec{\ell},
\end{equation}
which is independent of the path $\gamma_R-\gamma_L$ and has physical significance since it corresponds to the magnetic flux $\Phi_B$ inside the region enclosed by the charge. Then, the state of the system after it encircles the flux line is
\begin{equation}
    |\Psi\rangle = \frac{1}{\sqrt{2}}\left(|\psi_L^0\rangle + e^{i\Delta\phi_{AB}}|\psi_R^0\rangle\right), \label{shifted-state}
\end{equation}
where
\begin{equation}
    \Delta\phi_{AB} \equiv \frac{q}{\hbar} \oint \vec{A}\cdot d\vec{\ell}
    \label{AB effect}
\end{equation}
is the quantum phase accumulated by the charge, usually called the AB phase.

To verify that the AB phase is indeed a geometric phase, observe that, for a wavepacket $\Psi$ with center at $\vec{\ell}$ encircling the solenoid, the inner product $\langle\Psi(\vec{r}-\vec{\ell})|\vec{\nabla}_{\vec{\ell}}\Psi(\vec{r}-\vec{\ell})\rangle$ results in
\begin{equation}
    \int \Psi^*(\vec{r}-\vec{\ell}) \left[-i\frac{q}{\hbar} \vec{A}(\vec{\ell})+\vec{\nabla}_{\vec{\ell}}\right] \Psi(\vec{r}-\vec{\ell}) \ d^3r = -i \frac{q}{\hbar} \vec{A}(\vec{\ell}).
\end{equation}
Then, replacing it in Eq. \eqref{def-geom-phase},
\begin{equation}
    \phi_\text{geom} = i \oint_C \langle \Psi| \vec{\nabla}_{\vec{\ell}} \Psi\rangle\cdot d\vec{\ell} = \Delta\phi_{AB}.
    \label{eq-ref-geophase}
\end{equation}

Differently from various geometric phases, however, the AB phase does not depend on the path which $\vec{A}$ is integrated over, as already mentioned, which is one of the reasons it can be considered a \textit{topological phase}. Also, because there always exists a gauge in which vector potential vanishes in an arbitrary region that does not completely enclose the solenoid, \textit{a priori}, the AB effect cannot be seen as the result of the local interaction between the charge and the vector potential. As a result, it is typically considered to be of nonlocal nature. However, there is no final consensus on this issue and, thus, the origins of this phase are investigated until this day with works that include the idea of modular variables \cite{aharonov1969modular} and models that consider the source of magnetic field as part of the dynamics or quantize the magnetic field itself \cite{peshkin1961quantum, aharonov1991there, santos1999microscopic, choi2004exact, Vaidman2012, aharonov2015comment, vaidman2015reply, aharonov2016nonlocality, pearle2017quantized, li2018transition, marletto2020aharonov, saldanha2020shielded, horvat2020probing, paiva2021aharonov}. Furthermore, the interplay between Berry and AB phases in various scenarios can also be of interest. For instance, the phases acquired by a quantum charge with a large spreading while it encircles one or multiple solenoids were studied in Ref. \cite{aharonov1994aharonov}.

To conclude this section, we note that an electric AB effect was also proposed in the seminal article by Aharonov and Bohm \cite{Aharonov1959}. Moreover, a similar effect to the (magnetic) AB effect for neutral particles with a magnetic moment was later proposed by Aharonov and Casher \cite{aharonov1984topological}. This effect, often referred to as the Aharonov-Casher effect, was further studied and experimentally verified in various scenarios \cite{kaiser1988neutron, goldhaber1989comment, cimmino1989observation, dalibard2011colloquium}.

\section{Geometric phases and non-inertial effects}
\label{sec:sagnac-theory}

\begin{figure}
    \centering
    \includegraphics[width=8.6cm]{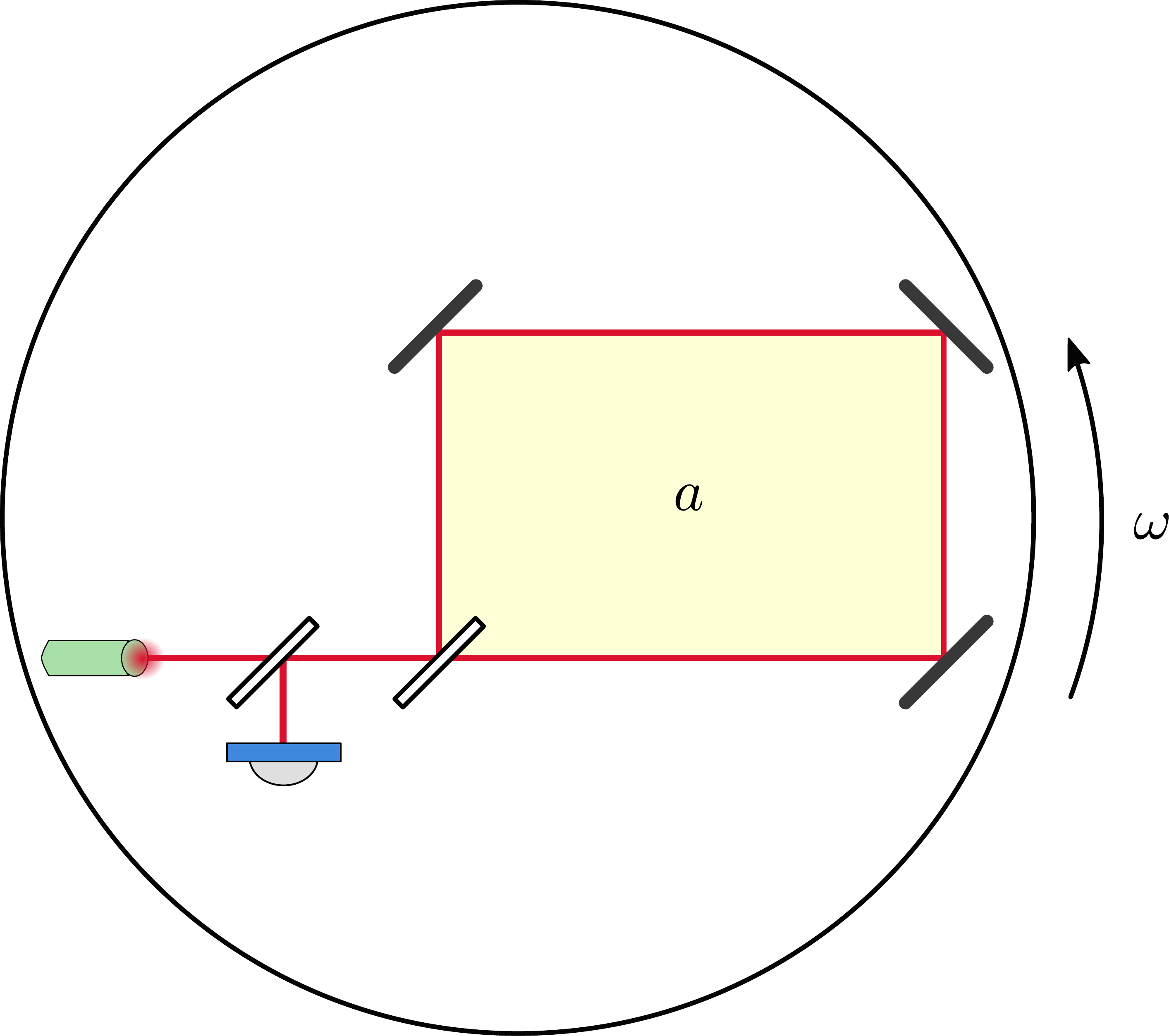}
    \caption{\textbf{Schematic representation of a standard Sagnac interferometer.} Two beams or wavepackets travel an interferometer in opposite directions. If the interferometer is rotating with angular speed $\omega$, the interference pattern is shifted by an amount that depends on $\omega$ and the area $a$ enclosed by the interferometer.}
    \label{fig:sagnac-generic}
\end{figure}

In 1913, Sagnac showed that angular rotation could be detected using interferometers in a classical setup \cite{sagnac1913preuve}. The configuration introduced by him is now known as the Sagnac interferometer. His result had been anticipated in two publications by Lodge \cite{lodge1893xv, lodge1897vi}, a prediction that, in turn, can be traced back to unpublished correspondences between Lodge and Larmor \cite{anderson1994sagnac}. This interferometer typically consists of two light beans enclosing a certain region with area $a$, which is rotating with angular speed $\omega$, as represented in Figure \ref{fig:sagnac-generic}. Because of the rotation of the region, the interference of the wavepackets is shifted by a phase
\begin{equation}
    \Delta\phi_S = \frac{8\pi a\omega}{c\lambda},
    \label{original-phase-s}
\end{equation}
where $\lambda$ is the initial wavelength of the photon. Based on this idea, in 1925, Michelson, Gale, and Pearson conducted an experiment to measure the effects of Earth's rotation on the speed of light \cite{michelson1925effect}.

One could, then, wonder whether such an effect had a quantum analog. This is indeed the case, as it was experimentally verified, for instance, by Collela, Overhauser, and Werner \cite{colella1975observation}, based on a previous theoretical proposal \cite{overhauser1974experimental}, and by Werner, Staudenmann, and Colella \cite{werner1979effect}. Nevertheless, it is possible to make distinctions between the ``classical'' and ``quantum'' Sagnac effects \cite{anandan1981sagnac, frauendiener2018notes, frauendiener2020gravitational}. In particular, in the nonrelativistic regime, there is no classical Sagnac effect, while the quantum version of it exists. However, in the relativistic limit, they become equivalent. Also, in the relativistic case, rotations of the interferometer or any other non-inertial influences are not required for the existence of a phase difference between the beams or wavepackets. In fact, because of the Doppler shift, the Sagnac effect can be obtained from a loop in space and relative motion --- even if both of the systems of reference involved are inertial \cite{tartaglia2015sagnac}.

The Sagnac effect is, of course, not exclusive to light-waves. In fact, it was observed in neutrons \cite{colella1975observation}, electrons \cite{hasselbach1993sagnac}, and atoms \cite{gustavson1997precision}. One may even consider a Sagnac effect for the superfluid Josephson interferometer \cite{anandan1981gravitational, *anandan1984gravitational}. This varied of systems is particularly relevant because, for precision purposes, we may benefit from the use of matter-waves since they have lower wavelengths, which increases the Sagnac phase-shift and the resulting sensitivity of the interferometer \cite{delgado2002quantum}. An analogous way to see this is by observing that the ratio between the rest mass $m$ of a particle and the effective mass $\hbar\omega/c^2$ of a light-wave with frequency $\omega$ amounts to a value with eleven orders of magnitude if the particle is an atom and the light-wave is an optical photon in the visible regime \cite{gustavson1997precision, barrett2014sagnac}. Then, if the Sagnac effect is seen as the result of the rotation of the interfering systems, matter-wave systems highly outperform light-wave systems in interferometers with equal areas. At the same time, optical systems have the advantage of higher particle fluxes and, typically, larger enclosed areas. Still, generally speaking, matter-wave systems are expected to outperform optical systems by several orders of magnitude \cite{gustavson1997precision}.

To see how the Sagnac effect is manifest for massive particles, we start from a classical analysis of a particle in a rotating system. Such a particle is subject to a Coriolis force given by
\begin{equation}
    \vec{F} = 2m \vec{u} \times \vec{\omega},
\end{equation}
where $\vec{u}$ indicates the velocity of the system (in the rotating system). Considering the potential vector $\vec{A}_r=\vec{\omega}\times\vec{r}$, we can write $\vec{w}=\vec{\nabla}\times \vec{A}_r$ and
\begin{equation}
    \vec{F} = 2m \vec{u} \times \left(\vec{\nabla}\times \vec{A}_r\right).
\end{equation}
From that, a Lagrangian can be defined, followed by its associated Hamiltonian
\begin{equation}
    H=\frac{1}{2m}\left(\vec{P}-2m\vec{A}_r\right)^2 = \frac{1}{2m}\left(\vec{P} - \frac{4\pi\hbar}{c\lambda} \vec{A}_r\right)^2.
\end{equation}
This Hamiltonian, in turn, can be quantized. In this case, it implies that, according to the direction of motion, a wavepacket acquires a phase of $\pm 4\pi a\omega/c\lambda$. Then, a system in a superposition of wavepackets moving in opposite directions ends up with a relative phase given by Eq. \eqref{original-phase-s}.

A more direct way to see this type of influence of non-inertial frames on the dynamics of quantum systems follows the reasoning presented in Ref. \cite{aharonov2014measure}. There, the authors showed that the dynamics of a free particle, i.e., a particle whose Hamiltonian is $H=(P_x^2+P_y)^2/2m$ in a rotating frame is described by a Hamiltonian with a term that is similar to a vector potential. In fact, if $\omega$ is the angular speed of the frame about the $z$ axis, the Hamiltonian $H'$ of the particle from the perspective of the rotating frame is a transformation of $H$ by the unitary $\exp(-i L_z \omega/\hbar)$, which in first order becomes
\begin{equation}
    H'\approx\frac{1}{2m} \left[\left(P_x+m\omega Y\right)^2+\left(P_y-m\omega X\right)^2\right]
    \label{hamil-ang-vel-alt}
\end{equation}
if the particle has an orbit with a well-defined radius. From Eqs. \eqref{hamil-ang-vel} and \eqref{hamil-ang-vel-alt}, we see that the rotation takes the usual place of the vector potential in the Hamiltonian.

From this discussion, we should expect a relation between the interference in the Sagnac interferometer and the AB effect. This is, in fact, the case, and it was presented in \cite{wang2004generalized}. Before discussing this result, we point out several differences between the two effects, e.g. while the AB effect is given in terms of $\hbar$ and vanishes in the classical limit, this is not the case for the Sagnac effect. In addition, the AB effect is topological while this effect is geometric.

By considering the setup with a rotating single mode optical fiber loop, the authors of Ref. \cite{wang2004generalized} first observed that the relative phase accumulated by each wavepacket is given by
\begin{equation}
    \Delta\phi_S = \frac{4\pi \vec{v}\cdot\Delta \vec{\ell}}{c\lambda},
\end{equation}
where $\vec{v}$ is the velocity of the fiber and $\Delta\ell$ is the length of a segment of the optical fiber. However, the important novelty of their work is that they also considered a more general scenario where only a portion of the circuit was moving. With that, they were able to gather evidence for the (expected) validity of the general expression for the phase $\Delta\phi_S$, which is
\begin{equation}
    \Delta\phi_S = \frac{4\pi}{c\lambda} \oint_\ell \vec{v}\cdot d\vec{\ell}.
\end{equation}
Comparing with the AB phase in Eq. \eqref{AB effect}, it can be seen that the parallel works with the mapping
\begin{equation}
    \vec{A} \rightarrow \frac{4\pi \hbar}{qc\lambda} \vec{v}.
\end{equation}
However, it should be noted that the vector potential and Berry connection are gauge-dependent quantities, while the angular velocity is not typically tied to a specific gauge. Still, this difference between these two quantities may not be as well defined as one may think if we consider that a choice of gauge is ultimately associated with a choice of frame \cite{aharonov1991there}. This aspect deserves to be further investigated since it is associated with a different discussion in the literature. On the one hand, in the nonrelativistic limit, the Sagnac effect can be seen as a manifestation of the lack of holonomy of the underlying geometry of the encircling particle, as we just showed --- and previously argued by, e.g., Chiao \cite{chiao1989berry} and Hendriks and Nienhuis \cite{hendricks1990sagnac}. As such, it is a Berry phase. On the other hand, this analogy between the Sagnac effect and AB effect or, more generally, Berry phases are not unanimous. For instance, while reviewing this effect, Malykin noted that Berry phases appear in addition to the Sagnac effect in different scenarios \cite{malykin2000sagnac}. Moreover, the AB and Berry phases are related to quantum systems, while the Sagnac effect can be attributed to classical systems. Because of it, Malykin concludes that, although valid, the analogy with AB and Berry phases is not of a fundamental nature. In any case, the interplay between different geometric phases in different implementations of the Sagnac effect seems worthy of further exploration.

Although the analysis presented in this section concerns the evolution of pure states, it can be, in principle, extended to the case of open and non-Hermitian systems, e.g., with some of the methods developed in Refs. \cite{garrison1988complex, zwanziger1991measuring, ning1992geometrical, bliokh1999appearance, sjoqvist2000geometric, singh2003geometric, filipp2003off, carollo2003geometric, tong2004kinematic, lombardo2006geometric, dietz2011exceptional}. This is still a subject of research that may shed light on many recent theoretical and technological developments. In particular, it could bring new insights into the role of rotating structures in studies of nonreciprocal optics and non-Hermitian physics, such as the works in Refs. \cite{huang2018nonreciprocal, zhang2020breaking, shi2021robust}.

In the next section, we present various applications of the Sagnac effect in quantum metrology and sensing, with special emphasis on the detection of gravitational effects.

\section{Applications of the Sagnac interferometer in quantum sensing}
\label{sec:sagnac-application}

In quantum sensing and metrology, one attempts to employ unique quantum properties such as single-particle interference, squeezing, and entanglement to improve classical techniques in terms of sensitivity, precision, or resolution. The probe particles could be massive or massless, but the goal in quantum metrology is typically to go beyond the shot-noise limit  (SNL), towards the Heisenberg limit. For achieving this goal, interferometers traversed by quantum states of light or matter are known to be invaluable tools \cite{giovannetti2004quantum, mitchell2004super, nagata2007beating, resch2007time, cooper2010entanglement, eberle2010quantum, wolf2019motional}.

For utilizing uniquely quantum states, various methods have been suggested and implemented. For instance, in the case of optical interferometers, the preferred method for this purpose is the employment of quantum squeezing \cite{caves1981quantum, xiao1987precision, schnabel2017squeezed} due to the simple generation of squeezed light by parametric amplification and the ability to seed a standard, high-power classical interferometer with squeezed light. This method claims usufruct rights for both the quantum advantage of squeezing together and the classical advantage of power \cite{ligo2011gravitational, barsotti2018squeezed, tse2019quantum}. However, all sub-shot-noise methods require a high detection efficiency, greater than 90\% for the SNL to be considerably overcome \cite{schnabel2017squeezed} and can be applied only in limited spectral ranges, where high-efficiency, low-noise detectors are available. 

The Sagnac interferometer is a major interferometric tool of high scientific and technological importance that can also be enhanced by squeezing. The Sagnac is the basis of optical and matter-wave gyroscopes, which are critical for military and civilian applications. In addition, the zero-area Sagnac interferometer was suggested \cite{traeger2000polarization,chen2003sagnac,eberle2010quantum} for the next generation of gravitational wave detectors.

\subsection{Quantum gyroscopes}

Since the Sagnac interferometer is sensitive to rotations, a natural application that takes advantage of this characteristic is its use in the construction of gyroscopes. As such, this type of application became common since they were first suggested or implemented in optical \cite{vali1976fiber, bergh1981all, bergh1981all2} and matter-wave \cite{riehle1991optical} setups. Assuming the G\"odel metric, they were even proposed as a device capable of placing an upper bound on the rotation of the universe \cite{delgado2002quantum}.

An important advantage of Sagnac interferometers over other configurations relies on their geometric nature. In fact, geometric phases depend exclusively on global variables and are, thus, intrinsically immune to local noise disturbances that preserve these geometric features \cite{carollo2003geometric, che2018phase}. Because of that, measurements of effects based on geometric phases can provide high-precision quantum sensing in real-world scenarios.

Here, we briefly present some advances in the development of Sagnac-based gyroscopes. To start, we focus on matter-wave gyroscopes. Since the first experimental application of the Sagnac effect with atoms \cite{riehle1991optical}, sensitivity improvements continued to be investigated \cite{gustavson1997precision, dowling1998correlated, gustavson2000rotation, wu2007demonstration}. In particular, the use of trapped atoms was suggested \cite{ketterle1992trapping, sauer2001storage, arnold2003adaptable} to overcome the complexity of implementing the atom gyroscopes \cite{moan2020quantum}. Because they support the atoms against gravity, the traps present longer measurement times without requiring large dropping distances, which otherwise would be challenging for matter-waves. With that, configurations using trapped atomic clocks \cite{stevenson2015, che2018phase} were shown to saturate the quantum Cram\'er-Rao bound. Moreover, entanglement-enhanced atomic gyroscopes with atoms trapped in an optical ring potential were proposed to achieve a precision of the order given by the Heisenberg limit \cite{cooper2010entanglement}.

On the optical side, it was first observed that single-mode fiber optic gyroscope had increased simplicity, stability, and the potential of very high rotation sensitivity \cite{vali1976fiber, bergh1981all, bergh1981all2}. However, the experiments began to really make use of quantum advantages after it was shown that the use of squeezed light in these gyroscopes could raise the sensitivity beyond the SNL \cite{marte1987enhanced, mehmet2010demonstration}. Entanglement was also shown to improve sensitivity in various schemes \cite{tian2018theoretical, fink2019entanglement, grace2020quantum}.

\subsection{Third generation of gravitational wave detectors}

Gravitational wave detectors have to meet harsh standards to function properly. For instance, in order to increase the information obtained from the measured interference pattern, they should use sophisticated feedback mechanisms allowing for major frequency and power stability. Moreover, they have to be extremely long to detect such small spacetime variations induced by gravitational waves. Also, like in most high-precision experiments with squeezed light, strict limits are imposed on the system's losses.

\begin{figure}
    \centering
    \includegraphics[width=\columnwidth]{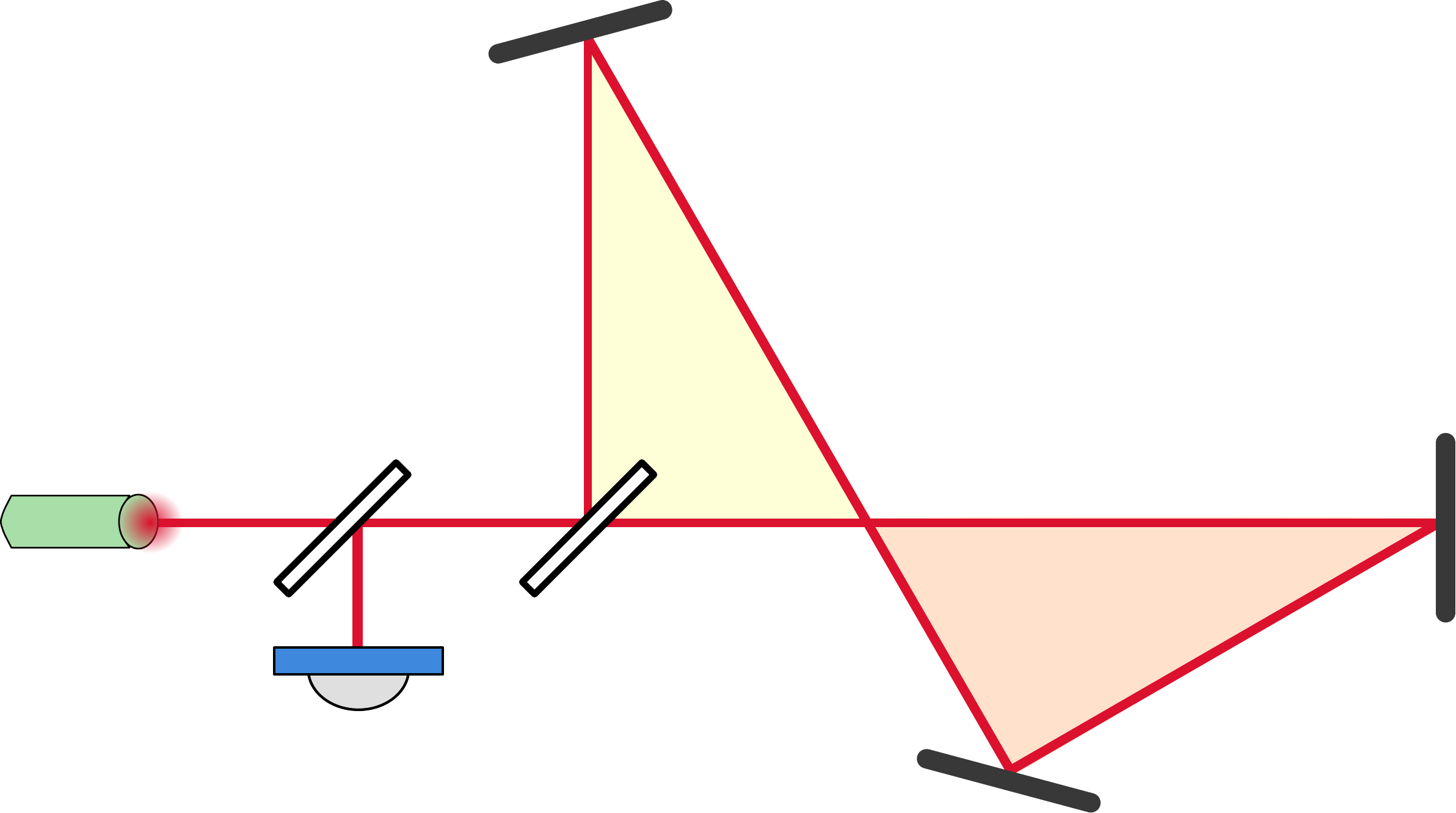}
    \caption{\textbf{Schematic representation of a simple zero-area Sagnac interferometer.} The disjoint oriented areas enclosed by systems travelling the interferometer (in yellow and orange) cancel each other. As a result, such an interferometer is insensitive to rotations.}
    \label{fig:circuit3}
\end{figure}

The Sagnac effect was suggested for detecting gravitational waves \cite{anandan1981sagnac}. With that and envisioning it as a possible replacement to the Michelson-Fabry-P\'erot-based LIGO, the zero-area Sagnac interferometer was introduced \cite{sun1996sagnac}. In this variation of the standard Sagnac interferometer, the waves still travel in opposite directions. However, the circuit is constructed in such a way to have area cancellation, as schematically illustrated in Figure \ref{fig:circuit3}. At first, it may seem that such an interferometer loses the most remarkable characteristic of the Sagnac interferometers: rotation sensitivity. However, some of the main advantages of this interferometer are its insensitivity to variations of laser frequency and its peak response in the frequency band of interest for LIGO applications. Moreover, this interferometer is also insensitive to mirror displacement at dc, thermally induced birefringence, and reflectivity imbalance in the arms. With all that, the optical tolerance requirements of the system are reduced and the system is more easily controlled.

Nevertheless, these advantages did not suffice to overcome some of the Sagnac topology disadvantages, like its low tolerance
to beamsplitter reflectivity error and beamsplitter tilt \cite{petrovichev1998simulating}. Furthermore, the ideal Sagnac interferometer did not present sensitivity advantages when compared to signal-recycled Michelson interferometers for the astrophysical needs of this type of application \cite{mizuno1997frequency}.

\begin{figure}
    \centering
    \includegraphics[width=\columnwidth]{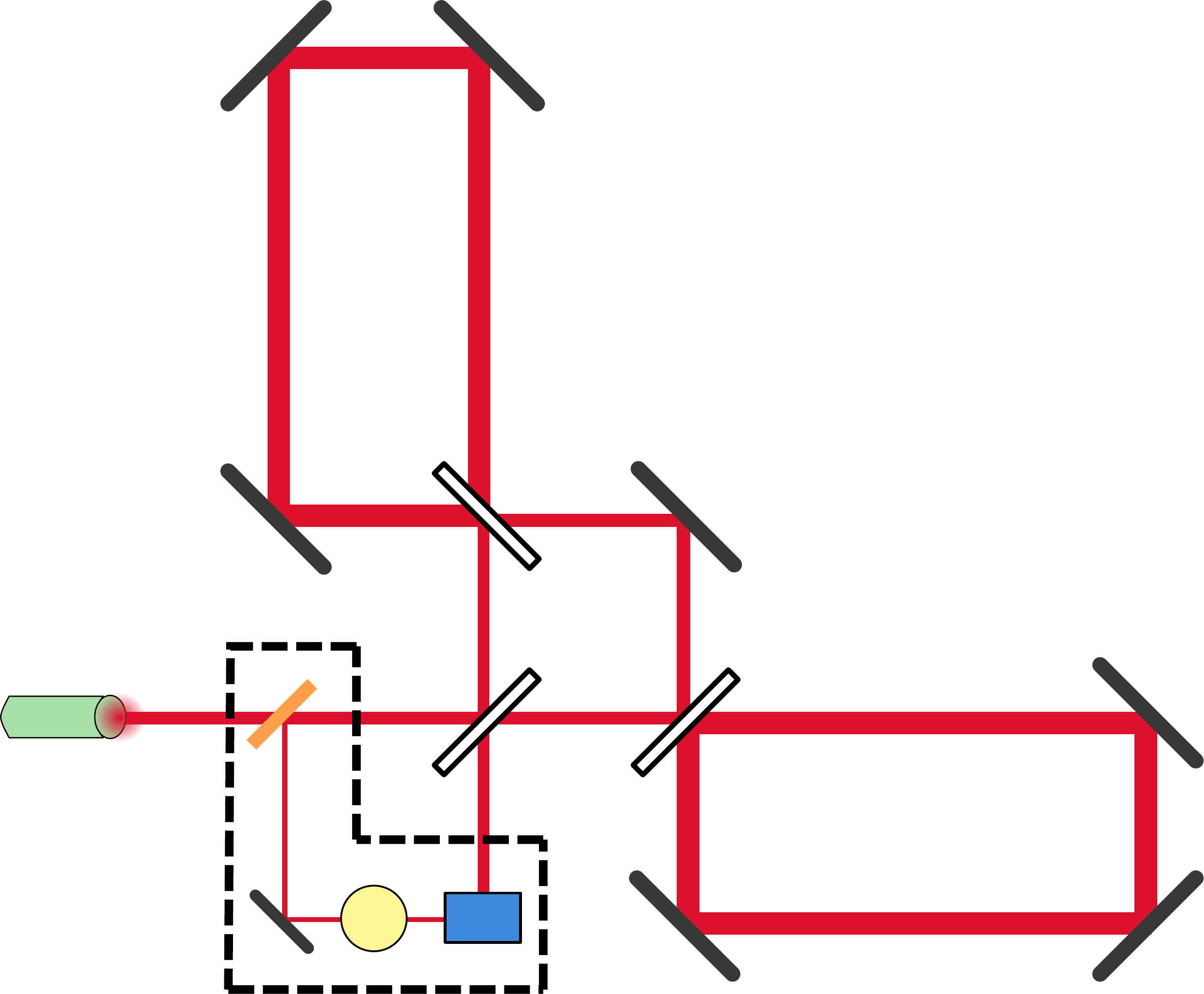}
    \caption{\textbf{Schematic representation of the ring-Sagnac interferometer.} The inner Sagnac interferometer, composed of the three white beamsplitters and a mirror (black element), is attached to two resonant cavities, one attached to its top left and the other to its bottom right. The elements inside the area delimited by the dashed line are used for homodyne readout. The orange beamsplitter separates a small portion of the signal. Then, this portion is reflected by a mirror and goes through a phase shifter (yellow circle). The blue box represents a homodyne detector. This setup was analysed in Ref. \cite{huttner2016candidates}.}
    \label{fig:circuit4}
\end{figure}

Despite these challenges, much effort was put into pushing the Sagnac scheme into an implementable state for the third generation of gravitational-wave detectors \cite{traeger2000polarization, chen2003sagnac, eberle2010quantum, voronchev2014sagnac, huttner2016candidates,huttner2020comparison}. For instance, it was shown that the zero-area Sagnac interferometer is a speedmeter, which can have advantages over position meters, like more conventional Michelson interferometers \cite{chen2003sagnac}. Moreover, the sensitivity of the zero-area Sagnac interferometer can be improved upon using ring cavities in the arms and signal recycling, similarly to the illustration in Figure \ref{fig:circuit4} without the elements inside the area delimited by the dashed line, which were a later addition. Although this would make the sensibility of this interferometer and the Michelson scheme comparable, the implementation of the Sagnac device had the advantage of not requiring the addition of any extra Fabry-P\'erot cavities.

Based on these results, it was suggested that further increase of the zero-area Sagnac interferometer's sensing could be achieved with the input of squeezed vacuum on its open port combined with a standard homodyne measurement \cite{eberle2010quantum}. However, such a measurement is inherently narrowband and requires near-ideal photodetectors as well as precise, technically demanding, active phase-locking of the squeezed vacuum to the local oscillator. Nevertheless, when compared to different interferometers, a Sagnac speedmeter is less susceptible to loss in a filter cavity \cite{voronchev2014sagnac}. This seems to firm the Sagnac topology as a good candidate for the third generation gravitational waves detectors since it simplifies the development of the filter cavity, reducing its implementation costs.

\begin{figure}
    \centering
    \includegraphics[width=\columnwidth]{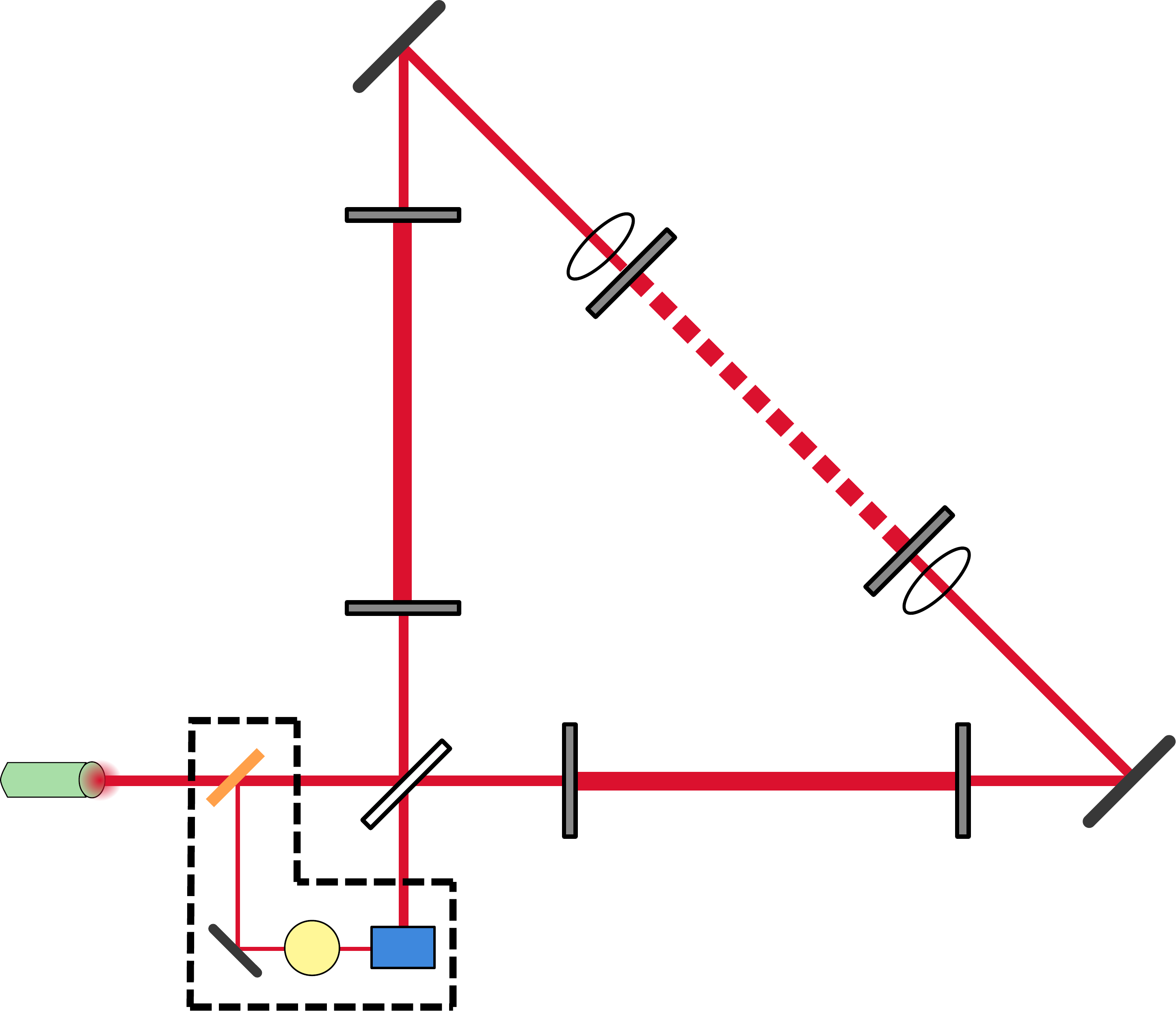}
    \caption{\textbf{Schematic representation of the sloshing-Sagnac interferometer.} The main part of the interferometer is composed of a beamsplitter (white element) and the horizontal and vertical arms. In each arm, there are two mirrors with equal reflectance and transmittance (gray elements). The signal leaked at the end of each of these cavities is reflected by a mirror (black element) and linked to an antiresonant cavity (in the diagonal). Lenses, represented by the elliptical elements, are used to match the cavity modes. Finally, the elements inside the area delimited by the dashed line are used for homodyne readout, like in the ring-Sagnac interferometer. This setup was investigated in Ref. \cite{huttner2016candidates}.}
    \label{fig:circuit5}
\end{figure}

More improvements were suggested with the introduction of a sloshing-Sagnac interferometer, which is made out of two resonant Fabry–P\'erot arm cavities in a Michelson configuration linked to a similar, antiresonant cavity running parallel to them \cite{huttner2016candidates}. This constitutes the Sagnac configuration represented in Figure \ref{fig:circuit5}. When compared to the ring-Sagnac interferometer shown in Figure \ref{fig:circuit4}, this device offers an extra degree of freedom for optimizations in the form of the finesse of the sloshing cavity that is separated from the arm cavity. In fact, with the use of squeezing, the sloshing-Sagnac interferometer presented a better performance in a lossy environment when compared to various other interferometers \cite{huttner2020comparison}.

Before concluding this subsection and proceeding to a broader class of sensing techniques, it should be noted that alternatives to Sagnac interferometers are also being considered in the literature. For instance, Michelson interferometers can also be adapted to become speedmeters, which may give them some advantages compared to the more conventional Michelson interferometers \cite{purdue2002analysis, purdue2002practical, freise2019simplified}.

\subsection{Enhanced sensing with weak value amplification}

Weak value amplification \cite{dixon2009ultrasensitive, susa2012optimal, dressel2013strengthening, jordan2014technical, pang2014entanglement, alves2015weak, harris2017weak} is a broad set of sensing techniques with diversified applications. It often consists of weakly coupling a quantum pointer to the quantum system of interest. The latter is pre- and post-selected, i.e., it has both initial and final boundary conditions $|\psi\rangle$ and $|\phi\rangle$, respectively. In this scenario, the shift of the pointer is, to first order in the coupling strength, proportional to the \textit{weak value} \cite{aharonov1988result}
\begin{equation}
    A_w=\frac{\langle \phi| A | \psi\rangle}{\langle \phi | \psi\rangle}
\end{equation}
of the measured operator $A$. As can be easily seen, the weak value of $A$ is not confined to its spectrum and can thus be very large. This amplification was shown to be particularly helpful in the presence of technical noise and detector saturation \cite{jordan2014technical, harris2017weak}. Related schemes use intermediate-strength or strong measurements for assessing weak values \cite{elitzur2011retrocausal,cohen2018determination, pan2020weak, dziewior2019universality} or, alternatively, rely on the inverse  weak values \cite{martinez2017ultrasensitive}. Weak values and weak measurements also bear interesting relations with the geometric phase \cite{sjoqvist2006geometric, cho2019emergence, gebhart2020topological, wang2021observing}.

Post-selection of the dark port of a Sagnac interferometer was shown to yield ultra-sensitive deflection \cite{dixon2009ultrasensitive} and tilt measurements \cite{martinez2017ultrasensitive}, as well as for sensitive gravimetry \cite{jordan2019gravitational}. Another work used weak value amplification for improved sensing of angular rotations \cite{magana2014amplification}. Enhancement of angular velocity measurements based on weak value amplification was recently demonstrated in Refs. \cite{fang2021weak, huang2021amplification}.

\section{Aharonov-Bohm effect and non-inertial systems}
\label{sec:ab-gravitational}

In this section, we discuss two effects that, to the best of our knowledge, have not been vastly considered in the literature and deserve more attention in practical applications.

From Larmor's theorem \cite{larmor1900aether}, it is known that a magnetic field $B$ acting on a particle can be emulated by the rotation of the frame with an angular speed proportional to $B$. It could be asked, then, if this also holds in the case of the AB effect since the particle only travels in regions with a null field. This question was answered in 1973 by Aharonov and Carmi \cite{aharonov1973quantum}, and their solution was further studied and employed in Refs. \cite{harris1980review, aharonov2014measure, aharonov2015comment}. They presented a direct analogy between the \textit{vector potential} and the angular velocity. As a result, this analogy allows the understanding of the AB phase geometrically.

\begin{figure}
	\centering
	\includegraphics[width=5cm]{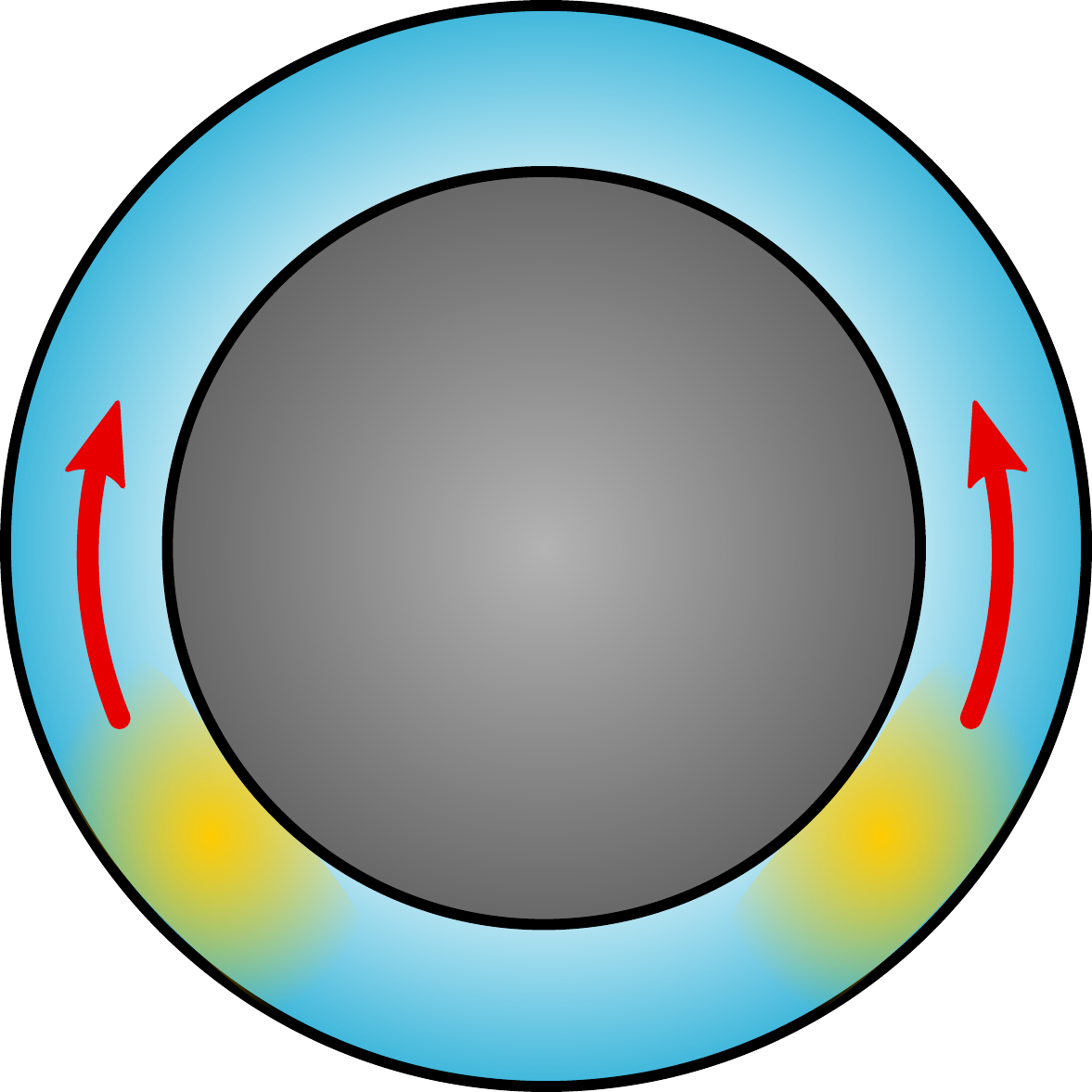}
	\caption{\textbf{Representation of a laboratory given by a narrow ring.} The massive disk (gray circle) and the laboratory (blue region) are assumed to rotate as a single system. Even if an electromagnetic field is used to counterbalance the gravitational field generated by the massive disk and the fictitious force created by its rotation inside the laboratory, a quantum experiment can still detect that the laboratory is not an inertial frame.}
	\label{fig6}
\end{figure}

To understand the result presented by Aharonov and Carmi, consider a lab given by a narrow ring with an inner radius $R_1$ and an outer radius $R_2$, as represented by the blue region in Figure \ref{fig6}. Also, assume that the ring rotates with an angular velocity $\omega$ and, for simplicity, that the ratio between the charge and mass is the same for every particle inside the ring. Moreover, the disc with radius $R_1$ is taken to be massive.

Because of its rotation, the ring experiences two pseudo-forces, namely the centrifugal and the Coriolis force. These forces, however, can be canceled by external electromagnetic forces. Indeed, the Coriolis force $\vec{F}_C$ acting on an object with mass $m$ and velocity $\vec{v}$ (measured in the lab) can be written as $\vec{F}_C=m\vec{v}\times\vec{C}$, where $\vec{C}$ is the field associated with $\vec{F}_C$. Moreover, because $\vec{C}$ satisfies $\vec{\nabla}\cdot\vec{C} = 0$, $\vec{F}_C$ is given by a field that is the rotation of a vector potential $\vec{A}_C$. Also, if $\vec{F}_c$ is the centrifugal force, $\vec{\nabla}\times\vec{F}_c =0$, i.e., $\vec{F}_c$ can be written as the gradient of a scalar potential.

Then, suppose an electromagnetic field is applied only \textit{inside} the ring to remove the pseudo-forces. Even in this case, a quantum experiment with a particle enclosing the disk $D_1$ (with radius $R_1$) in a superposition of wavepackets traveling in different paths, as represented in Figure \ref{fig6}, can detect that the ring is not an inertial frame. In fact, denoting by $\vec{A}_T$ the vector potential associated with the rotating mass after the inclusion of the magnetic field in the lab, it is possible to write the Hamiltonian of the system as
\begin{equation}
    H = \frac{1}{2m} \left(\vec{P}-m\vec{A}_T\right)^2,
    \label{hamil-ang-vel}
\end{equation}
which implies that the relative phase accumulated by the wavepackets is proportional to
\begin{equation}
    \Delta\phi_R = m\oint \vec{A}_T \cdot d\vec{\ell} = m\int_{D_1} \vec{C} \cdot d\vec{S} = 2\pi m R_1^2 \omega,
\end{equation}
i.e., the accumulated phase is proportional to the angular speed $\omega$ of the lab. The Hamiltonian in Eq. \eqref{hamil-ang-vel} compares to the Hamiltonian associated with the AB effect in Eq. \eqref{eq-ham-vp}. Moreover, a computation similar to the one performed in Eq. \eqref{eq-ref-geophase} shows that $\Delta\phi_R$ is indeed a geometric phase.

This phenomenon is an AB-like effect. It suggests the necessity for a modification of the equivalence principle in quantum theories. In fact, the charge travels in regions where the effective field is null, although its potential is not. In particular, we can deduce from the previous example that, in a general relativistic treatment, curvature effects, and not just accelerations caused by them, can be detected in quantum interference experiments. Indeed, low-order curvature effects were studied in an interferometer with thermal neutrons that only makes use of horizontal mirrors \cite{anandan1984curvature}.

Moreover, another similarity between the example studied in this section and the magnetic AB effect is that the trajectories of the wavepackets can be deformed and, as long as they do not enter the gray region in Figure \ref{fig6}, the acquired relative phase between them is unchanged. However, differently from the magnetic AB effect, the phase considered here is acquired in a manifestly local manner.

One can go even further and discuss the gravitational AB effect \cite{anandan1977gravitational, anandan1979interference, anandan1983interferometry, ford81, chiao2014gravitational}. In particular, the authors of Ref. \cite{chiao2014gravitational} replace the magnetic flux with a Lense-Thirring field. Then, they show a relation between this effect and gravitational radiation and parametric oscillators. Generally speaking, it seems that the gravitational AB effect may enjoy the better sensitivity of the proposed squeezing-enhanced Sagnac interferometer. However, this is a practical question that deserves further exploration.

\section{Discussion and outlook}
\label{sec:discussion}

The geometry of quantum states is remarkable in breadth and depth. We have only described a small selection of its various fundamental and practical merits in this review article. On the fundamental side, it arises from a rich mathematical structure that can be related to a classical counterpart via the Bohr-Sommerfeld quantization rule, as we explained briefly in Section \ref{sec:geom-phases}. Moreover, AB-like non-inertial and gravitational effects introduce interesting quantum phenomena and even suggest the necessity of a reformulation of basic concepts, like the equivalence principle, as seen in Section \ref{sec:ab-gravitational}. On the application side, particular attention was given to gravitational and non-inertial measurements. However, geometric phases are also an important player in quantum information and computation \cite{vedral2003geometric, sjoqvist2015geometric, chen2020observable}, chemical physics \cite{zwanziger1990berry, mead1992geometric, kuppermann1993geometric, kendrick2015geometric}, and many other areas.

Before concluding, we outline some topics for further research. From a practical perspective, the geometric and gravitational AB effects do not seem to be vastly studied in the literature and may lead to new quantum-enhanced precision measurements. These effects and their relation to other relativistic ones, such as frame-dragging, seem to deserve further attention.

On the fundamental level, there are still some open questions such as: Does the analogy between various geometric phases and the Sagnac effect represent a genuine physical relation between them? Is the AB effect local? If so, in which sense? Can constructions of a complete quantum mechanical description of systems which does not utilize potentials help in this investigation?

Regarding this last question, it indeed seems that an interesting direction is the study of fully quantized systems where gauge-dependence is linked to frame-dependence \cite{aharonov1991there, paiva2021aharonov}. This approach could reveal similarities and also highlight important differences between the AB and other geometric phases.

In this approach and other similar ones, geometric and topological phases were linked to the creation of entanglement. For instance, in the case of the AB effect, if the source of the magnetic field is not an eigenstate of a relevant observable, the source and the charge encircling it become entangled \cite{paiva2021aharonov}. Furthermore, as argued by Vaidman \cite{Vaidman2012}, it is possible to conceive a model in which entanglement is present even if initially there is no uncertainty in the flux. This suggests that a general interplay may exist between dynamical nonlocality and kinematical nonlocality (see also \cite{marletto2020aharonov}). The latter is a broad category that includes, for instance, Bell scenarios. The former is a type of nonlocality that emerges from the dynamical equations of motion  \cite{aharonov1969modular, aharonov2005quantum, aharonov2017finally} and can be associated, e.g., with the AB effect. Then, a better understanding of how entanglement is related to these phases could have fundamental and practical consequences.

Another research question that may be worth considering concerns the classical study of an electron moving in a loop next to a solenoid with a dynamical source of magnetic field. Can the interaction between the field and the electron be interpreted in terms of geometric phases? Since the AB effect is broadly perceived as a genuine quantum effect, the expected answer is, in principle, negative. However, in light of the geometric analogs discussed above the analysis might be more subtle. At least in a classical treatment where an electromagnetic wave interacts with the potential generated by a solenoid source, Vedral recently argued in favor of the existence of a classical account of the effect \cite{vedral2021classical}.

Finally, one may also study to what extent other effects associated with geometric phases, like the optical Magnus effect \cite{bolotovskii1977optical, zel1990rotation, dooghin1992optical, bliokh2004topological, bliokh2004modified}, are fundamentally related or can benefit from the Sagnac effect.

\acknowledgements{We thank Ady Arie, Avi Pe'er, and Michael Rosenbluh for many helpful discussions. This research was supported by grant number FQXi-RFP-CPW-2006 from the Foundational Questions Institute and Fetzer Franklin Fund, a donor-advised fund of Silicon Valley Community Foundation, by the Israeli Innovation authority (grants 70002, 73795), by the Pazy foundation, by the Israeli Ministry of Science and Technology and by the Quantum Science and Technology Program of the Israeli Council of Higher Education.}

\section*{Conflict of Interest}
The authors declare no conflict of interest.

\section*{Key Words}
Geometric phase, Sagnac effect, Aharonov-Bohm effect

\bibliography{references}

\end{document}